\documentclass[a4paper,twocolumn,english,aps,pra,superscriptaddress,floatfix,showpacs]{revtex4}
\usepackage[T1]{fontenc}
\usepackage[latin9]{inputenc}
\usepackage{amsmath}
\usepackage{amssymb}
\usepackage{esint}

\makeatletter

\@ifundefined{textcolor}{}
{%
 \definecolor{BLACK}{gray}{0}
 \definecolor{WHITE}{gray}{1}
 \definecolor{RED}{rgb}{1,0,0}
 \definecolor{GREEN}{rgb}{0,1,0}
 \definecolor{BLUE}{rgb}{0,0,1}
 \definecolor{CYAN}{cmyk}{1,0,0,0}
 \definecolor{MAGENTA}{cmyk}{0,1,0,0}
 \definecolor{YELLOW}{cmyk}{0,0,1,0}
 }

\makeatother

\usepackage{babel}
\begin{document}

\title{Geometric Phases for Coherent States }

\author{Da-Bao Yang }

\affiliation{Theoretical Physics Division, Chern Institute of Mathematics, Nankai
University, Tianjin 300071, People's Republic of China}
\affiliation{Centre for Quantum Technologies, National
University of Singapore, 3 Science Drive 2, Singapore 117543}

\author{Jing-Ling Chen}

\email{chenjl@nankai.edu.cn}

\selectlanguage{english}%

\affiliation{Theoretical Physics Division, Chern Institute of Mathematics, Nankai
University, Tianjin 300071, People's Republic of China}
\affiliation{Centre for Quantum Technologies, National
University of Singapore, 3 Science Drive 2, Singapore 117543}

\author{Chunfeng Wu}

\affiliation{Centre for Quantum Technologies, National
University of Singapore, 3 Science Drive 2, Singapore 117543}

\author{C. H. Oh}

\email{phyohch@nus.edu.sg}

\selectlanguage{english}%

\affiliation{Centre for Quantum Technologies, National
University of Singapore, 3 Science Drive 2, Singapore 117543}
\affiliation{Department of Physics, National
University of Singapore, 2 Science Drive 3, Singapore 117542}

\date{\today}
\begin{abstract}
We explore geometric phases of coherent states and some of their properties.
A better and elegant expression of geometric phase for coherent
state is derived. It is used to obtain the explicit form of the geometric
phase for entangled coherent states, several interesting results followed
by considering different cases for the parameters. The effects of
entanglement and harmonic potential on the geometric phase are discussed.

\end{abstract}


\maketitle

\section{Introduction}

\label{sec:introduction} The phase factor of a wave function is one
of the most fundamental concepts in quantum physics. It is Berry who
first discovered that a geometric phase can be accrued in the wave
function of a quantum system in adiabatic, unitary, and cyclic evolution
of time-dependent quantum system \cite{berry1984quantal}. Three years
later, by abandoning the context of adiabaticity, Berry's result was
extended by Aharonov and Anandan \cite{aharonov1987phase}. Soonafter
Samuel and Bhandari \cite{samuel1988general} further generalized
the Aharonov and Anandan phase to non-cyclic and non-unitary cases
by resorting to Pancharatnam's pioneer work \cite{pancharatnam1956connection}.
Subsequently, Ref. \cite{mukunda1993quantum} established the quantum
kinematic approach to geometric phases, which is the most general
theory on geometric phases of pure quantum states. However the above
definitions of geometric phases are not applicable, if the initial
and final states are orthogonal. Manini and Pistolesi \cite{manini2000off}
first proposed a method to solve the problem by introducing the Abelian
off-diagonal geometric phases during adiabatic evolution. One year
later the definition was generalized to nonadiabatic cases with the
use of Bargmann invariants \cite{mukunda2001bargmann}. Recently,
Kult \emph{et al}. \cite{kult2007nonabelian} made a step forward
in this direction by extending the concepts to non-Abelian systems.
Despite all the results achieved for pure states in the literatures,
the definitions of geometric phases have also been generalized to
the case of mixed states. Uhlmann \cite{uhlmann1986parallel} presented
a definition in mathematical context of purification. Sj{ö}qvist
\emph{et al}. \cite{sjoqvist2000mixed} developed the non-degenerate
geometric phase in non-cyclic and unitary evolution under the background
of quantum interference, which is independent of surroundings. Extensions
of mixed-state geometric phases to the degenerate case \cite{singh2003geometric}
and the kinematic approach \cite{tong2004kinematic} have also been
achieved.

The Berry phase and its generalizations have given rise to many applications
ranging from condensed matter physics \cite{gpa1} to quantum information
science \cite{gpa2,gpa3,gpa4,gpa5}. For example, geometric phase
plays an important role in quantum information and computation protocols
which employ dynamical or geometric phases to achieve quantum gates.
Dynamical phase gates require a precise control of the pulse area.
Geometric phases depend only on the solid angle enclosed by the parameter
path and generally not on the details of the path. Thus geometric
phases can render robust protocols for quantum computation \cite{gpa3,gpa4,gpa5}.
The promising applications spawn various theoretical investigations
on geometric phases of different physical systems \cite{Sjoqvist2000geometric,tong2003entangled,yang2011aharonov,chaturvedi1987berry,pati1995geometric,yang2011geometric}.
Ref. \cite{Sjoqvist2000geometric} reported entanglement dependence
of the non-cyclic and non-adiabatic geometric phase for entangled spin pairs
in a static magnetic field, and the result was promoted to entangled
spin particles in a rotating magnetic field later \cite{tong2003entangled}.
Literatures \cite{chaturvedi1987berry,pati1995geometric,yang2011geometric}
explored the geometric phases of coherent states of a one-dimensional
harmonic oscillator. Chaturvedi \textit{et al.} found the Berry phase
for coherent states \cite{chaturvedi1987berry}; the geometric phase
for the noncyclic evolution of coherent states was studied in Ref.
\cite{pati1995geometric}; and the authors in \cite{yang2011geometric}
formulated the non-unitary and non-cyclic geometric phases for nonlinear
coherent states. 

In this work we study the geometric phases of coherent states, especially
entangled coherent states. In the next section, we briefly review
some concepts of kinematic approach to geometric phases. These ideas
are illustrated by considering the geometric phase for coherent states
of a one-dimensional harmonic oscillator. We find an equivalent but
more elegant form of geometric phase than that given in Ref. \cite{pati1995geometric}.
In Sec. III, the geometric phases of entangled coherent states are
calculated and some of their properties are discussed. We study the
influences of entanglement and harmonic potential on the geometric
phases. We end with conclusion in the last section.

\section{Brief Review of Quantum Kinematic Approach to Geometric Phase}

\label{sec:reviews}

In this section, the rudiment about the kinematic approach to the
non-cyclic geometric phase \cite{mukunda1993quantum} is reviewed.
When a quantum system undergoes a unitary evolution, its state vectors
trace a smooth trajectory in Hilbert space, which is $\mathcal{C}=\{|\psi(t)\rangle\in\mathcal{H}|t\in[0,\tau]\subset\mathcal{R}\}$.
Accompanying the quantum evolution, there exists a geometric quantity
that is both gauge invariant and reparametrization invariant, under
the condition that $\langle\psi(0)|\psi(\tau)\rangle\neq0$. Such
a geometric quantity is called geometric phase, which is expressed
as follows
\begin{equation}
\gamma=\chi-\delta,\label{eq:GeometricPhase}
\end{equation}
 where $\chi$ is total phase
\begin{equation}
\chi={\rm arg}(\langle\psi(0)|\psi(\tau)\rangle),\label{eq:TotalPhase}
\end{equation}
 and $\delta$ is dynamical phase
\begin{equation}
\delta=-i\int_{t=0}^{t=\tau}\langle\psi(t)|\frac{d}{dt}|\psi(t)\rangle.\label{eq:DynamicalPhase}
\end{equation}

In order to illustrate the above definition clearly, we consider a
one-dimensional harmonic oscillator whose Hamiltonian reads
\[
H=\frac{p^{2}}{2m}+\frac{1}{2}kx^{2},
\]
 and the state vector at any later time $\tau$ is given by
\begin{equation}
|\alpha,\tau\rangle=e^{-|\alpha|^{2}/2}\sum_{n=0}^{\infty}\frac{\alpha^{n}}{\sqrt{n!}}e^{-i\omega\tau(n+1/2)}|n\rangle.\label{eq:CoherentState}
\end{equation}
 where $\alpha$ is a complex number representing the coherent state
and $n$ labels number state. According to Eq. \eqref{eq:TotalPhase},
the total phase is
\begin{equation}
\chi={\rm arg}(\langle\alpha,0|\alpha,\tau\rangle)=-(|\alpha|^{2}\sin\omega\tau+\frac{1}{2}\omega\tau),\label{eq:TotalPhaseOfOneParticle}
\end{equation}
 where we have used the inner product $\langle\alpha,0|\alpha,\tau\rangle=e^{-|\alpha|^{2}(1-\cos\omega\tau)}e^{-i(|\alpha|^{2}\sin\omega\tau+\frac{1}{2}\omega\tau)}$.
By using Eq. \eqref{eq:DynamicalPhase}, the dynamical phase is equal
to
\begin{equation}
\delta(\alpha)=-\omega\tau(\frac{1}{2}+|\alpha|^{2}).\label{eq:DyanmicalPhaseOfOneParticle}
\end{equation}
 Substituting Eq. \eqref{eq:TotalPhaseOfOneParticle} and \eqref{eq:DyanmicalPhaseOfOneParticle}
into Eq. \eqref{eq:GeometricPhase}, one obtains the corresponding
geometric phase that takes the form
\begin{equation}
\gamma=|\alpha|^{2}(\omega\tau-\sin\omega\tau).\label{eq:GeometricPhaseOfOneParticle}
\end{equation}
 Let us point out that the geometric phase \eqref{eq:GeometricPhaseOfOneParticle}
is an elegant result than that of Ref. \cite{pati1995geometric}.
When $\omega\tau=2\pi$, the known cyclic geometric phase
\begin{equation}
\gamma=|\alpha|^{2}\omega\tau
\end{equation}
 is recovered \cite{pati1995geometric}.

\section{Geometric Phases for Entangled Coherent States}

\label{sec:EntangledCoherentStates}

\subsection{Entangled coherent states of harmonic oscillators}

$\:$Pertaining to harmonic oscillators, a general unnormalized two-particle
entangled coherent state \cite{wielinga1992entangled} is of the form
\begin{equation}
|\Psi(t)\rangle=e^{-i\varphi/2}\cos\frac{\theta}{2}|\alpha,t\rangle_{1}|\mu,t\rangle_{2}+e^{i\varphi/2}\sin\frac{\theta}{2}|\beta,t\rangle_{1}|\nu,t\rangle_{2},\label{eq:EntangledCoherentState}
\end{equation}
 where $\alpha$, $\beta$, $\mu$ and $\nu$ are complex parameters
of the corresponding coherent states \eqref{eq:CoherentState}, the
subscripts denote particle 1 and 2 respectively, and $\theta$ as
well as $\varphi$ are real numbers. $\theta$ is a quantity which
determines the degree of entanglement of the two-particle entangled
coherent state. When $\theta=0$ or $\theta=\pi$, the entangled coherent
state \eqref{eq:EntangledCoherentState} reduces to product state;
when $\theta=\pi/2$, the degree of entangled coherent state becomes
maximal. Note that the coherent state vectors of subsystems are generally
not orthogonal to each other:
\begin{eqnarray}
\langle\alpha,t|\beta,t\rangle=\exp[-\frac{1}{2}(|\alpha|^{2}+|\beta|^{2})+\alpha^{*}\beta],
\end{eqnarray}
 which is quite different from the case of entangled spin-1/2 pairs
\cite{Sjoqvist2000geometric}. The evolutionary state studied in \cite{Sjoqvist2000geometric}
is given by
\begin{eqnarray}
|\Psi(t)\rangle & = & e^{-i\varphi/2}\cos\frac{\theta}{2}|+\textbf{n},t\rangle_{1}|+\textbf{m},t\rangle_{2}\nonumber \\
 &  & +e^{i\varphi/2}\sin\frac{\theta}{2}|-\textbf{n},t\rangle_{1}|-\textbf{m},t\rangle_{2},
\end{eqnarray}
 where $|+\textbf{n},t\rangle_{1}$ (or $|+\textbf{m},t\rangle_{2}$)and
$|-\textbf{n},t\rangle_{1}$ (or $|-\textbf{m},t\rangle_{2}$) are
mutually orthogonal. Hence it is reasonable to expect that the total
phase, dynamical phase and geometric phase of entangled coherent states
will exhibit some different features from that of entangled spin pairs
as given in Ref. \cite{Sjoqvist2000geometric}. For the sake of completeness,
we write down the normalized entangled coherent state in the following
\begin{eqnarray}
|\psi(t)\rangle & = & \frac{1}{\mathcal{N}}(e^{-i\varphi/2}\cos\frac{\theta}{2}|\alpha,t\rangle_{1}|\mu,t\rangle_{2}\nonumber \\
 &  & +e^{i\varphi/2}\sin\frac{\theta}{2}|\beta,t\rangle_{1}|\nu,t\rangle_{2}),\label{eq:NormalizedEntangledCoherentState}
\end{eqnarray}
 where the normalization factor satisfies
\begin{eqnarray}
\mathcal{N}^{2} & = & 1+\sin\theta\exp[-\frac{1}{2}(|\alpha|^{2}+|\beta|^{2})-\frac{1}{2}(|\mu|^{2}+|\nu|^{2})]\times\nonumber \\
 &  & {\rm Re}[\exp(i\varphi+\alpha^{*}\beta+\mu^{*}\nu)],
\end{eqnarray}
 which is independent of time and for simplicity we choose $\mathcal{N}$
as a positive real number.

\subsection{Geometric phases for entangled coherent states}

Consider a system that consists of two non-interacting entangled coherent
particles which are in harmonic potentials. The corresponding Hamiltonian
reads
\begin{eqnarray}
H=\frac{p_{1}^{2}}{2m_{1}}+\frac{1}{2}k_{1}x_{1}^{2}+\frac{p_{2}^{2}}{2m_{2}}+\frac{1}{2}k_{2}x_{2}^{2}.
\end{eqnarray}
 Because there is not interaction between the two particles, its time
evolution operator takes a product form. If we choose initial state
as
\begin{eqnarray}
|\psi(0)\rangle & = & \frac{1}{\mathcal{N}}(e^{-i\varphi/2}\cos\frac{\theta}{2}|\alpha,0\rangle_{1}|\mu,0\rangle_{2}\nonumber \\
 &  & +e^{i\varphi/2}\sin\frac{\theta}{2}|\beta,0\rangle_{1}|\nu,0\rangle_{2}),
\end{eqnarray}
 then at any later time $t>0$, the state vector can be written in
the form as shown in Eq. \eqref{eq:NormalizedEntangledCoherentState}.

In order to calculate the geometric phases of this system, we need
to determine its total phase and dynamical phase respectively. By
the use of the formula below
\begin{eqnarray}
\langle\alpha,0|\beta,\tau\rangle=\exp[-\frac{1}{2}(|\alpha|^{2}+|\beta|^{2})+\alpha^{*}\beta e^{-i\omega\tau}-i\frac{1}{2}\omega\tau],
\end{eqnarray}
 we get \begin{widetext}
\begin{eqnarray}
\begin{array}{rcl}
\mathcal{N}^{2}\langle\psi(0)|\psi(\tau)\rangle & = & \frac{1+\cos\theta}{2}f(\alpha;\mu)\exp[i\chi(\alpha;\mu)]+\frac{1-\cos\theta}{2}f(\beta;\nu)\exp[i\chi(\beta;\nu)]\\
 &  & +\frac{1}{2}\sin\theta g(\alpha,\beta;\mu,\nu)\exp\{i[h(\alpha,\beta;\mu,\nu)+\varphi]\}+\frac{1}{2}\sin\theta g(\beta,\alpha;\mu,\nu)\exp\{i[h(\beta,\alpha;\mu,\nu)-\varphi]\},
\end{array}
\end{eqnarray}
 where
\begin{eqnarray}
\begin{array}{rcl}
f(\alpha;\mu) & = & \exp[-\rho_{\alpha}^{2}(1-\cos\omega_{1}\tau)-\rho_{\mu}^{2}(1-\cos\omega_{2}\tau)],\\
g(\alpha,\beta;\mu,\nu) & = & \exp[-\frac{1}{2}(\rho_{\alpha}^{2}+\rho_{\beta}^{2})-\frac{1}{2}(\rho_{\mu}^{2}+\rho_{\nu}^{2})+\rho_{\alpha}\rho_{\beta}\cos(\phi_{\alpha}-\phi_{\beta}+\omega_{1}\tau)+\rho_{\mu}\rho_{\nu}\cos(\phi_{\mu}-\phi_{\nu}+\omega_{2}\tau)],\\
\chi(\alpha;\mu) & = & -(\rho_{\alpha}^{2}\sin\omega_{1}\tau+\frac{1}{2}\omega_{1}\tau)-(\rho_{\mu}^{2}\sin\omega_{2}\tau+\frac{1}{2}\omega_{2}\tau),\\
h(\alpha,\beta;\mu,\nu) & = & -[\rho_{\alpha}\rho_{\beta}\sin(\phi_{\alpha}-\phi_{\beta}+\omega_{1}\tau)+\frac{1}{2}\omega_{1}\tau]-[\rho_{\mu}\rho_{\nu}\sin(\phi_{\mu}-\phi_{\nu}+\omega_{2}\tau)+\frac{1}{2}\omega_{2}\tau],
\end{array}
\end{eqnarray}
 with $\rho_{\lambda}$ and $\phi_{\lambda}$ being the amplitude
and phase of the state $|\lambda\rangle$ and $\lambda=\rho_{\lambda}e^{i\phi_{\lambda}}\;\;(\lambda=\alpha,\beta,\mu,\nu)$.
Hence, the total phase is found to be
\begin{equation}
\chi(\alpha,\beta;\mu,\nu)=\arg(\langle\psi(0)|\psi(\tau)\rangle)=\arctan\frac{A}{B},\label{eq:TotalPhaseOfEntangledCoherentStates}
\end{equation}
 where
\begin{eqnarray}
\begin{array}{ccl}
A & = & {\rm Im}(\langle\psi(0)|\psi(\tau)\rangle)\\
 & = & (1+\cos\theta)f(\alpha;\mu)\sin[\chi(\alpha;\mu)]+(1-\cos\theta)f(\beta;\nu)\sin[\chi(\beta;\nu)]\\
 &  & +\sin\theta g(\alpha,\beta;\mu,\nu)\sin[h(\alpha,\beta;\mu,\nu)+\varphi]+\sin\theta g(\beta,\alpha;\nu,\mu)\sin[h(\beta,\alpha;\nu,\mu)-\varphi],
\end{array}
\end{eqnarray}
 and
\begin{eqnarray}
\begin{array}{ccl}
B & = & {\rm Re}(\langle\psi(0)|\psi(\tau)\rangle)\\
 & = & (1+\cos\theta)f(\alpha;\mu)\cos[\chi(\alpha;\mu)]+(1-\cos\theta)f(\beta;\nu)\cos[\chi(\beta;\nu)]\\
 &  & +\sin\theta g(\alpha,\beta;\mu,\nu)\cos[h(\alpha,\beta;\mu,\nu)+\varphi]+\sin\theta g(\beta,\alpha;\nu,\mu)\cos[h(\beta,\alpha;\nu,\mu)-\varphi].
\end{array}
\end{eqnarray}
 In the following, the dynamical phase is determined by using the
formula
\begin{eqnarray}
\langle\alpha,t|\frac{d}{dt}|\beta,t\rangle=-i\omega\exp[-\frac{1}{2}(|\alpha|^{2}+|\beta|^{2})-\alpha^{*}\beta](\frac{1}{2}+\alpha^{*}\beta).
\end{eqnarray}
 The dynamical phase \eqref{eq:DynamicalPhase} is then found to be
\begin{equation}
\begin{array}{ccl}
\delta(\alpha,\beta;\mu,\nu) & = & -\frac{1}{\mathcal{N}^{2}}\frac{1+\cos\theta}{2}[\omega_{1}\tau(\frac{1}{2}+|\alpha|^{2})+\omega_{2}\tau(\frac{1}{2}+|\mu|^{2})]-\frac{1}{\mathcal{N}^{2}}\frac{1-\cos\theta}{2}[\omega_{1}\tau(\frac{1}{2}+|\beta|^{2})\\
 &  & +\omega_{2}\tau(\frac{1}{2}+|\nu|^{2})]-\frac{1}{\mathcal{N}^{2}}{\rm Re}\{\sin\theta e^{i\varphi}\exp[-\frac{1}{2}(|\alpha|^{2}+|\beta|^{2})+\alpha^{*}\beta\\
 &  & -\frac{1}{2}(|\mu|^{2}+|\nu|^{2})+\mu^{*}\nu][\omega_{1}\tau(\frac{1}{2}+\alpha^{*}\beta)+\omega_{2}\tau(\frac{1}{2}+\mu^{*}\nu)]\}
\end{array}.\label{eq:DynamicalPhaseOfEntangledCoherentState}
\end{equation}
 \end{widetext} 

According to Eq. \eqref{eq:GeometricPhase}, we can obtain the geometric
phase as
\begin{equation}
\gamma(\alpha,\beta;\mu,\nu)=\chi(\alpha,\beta;\mu,\nu)-\delta(\alpha,\beta;\mu,\nu).\label{eq:GeometricPhaseOfEntangledCoherentState}
\end{equation}
 However, the above expression is too tedious to use. So we concentrate
on a particular type of entangled coherent states by setting $\beta=-\alpha$
and $\nu=-\mu$, and thus the geometric phase reads \begin{widetext}
\begin{eqnarray}
 &  & \gamma(\alpha,-\alpha;\mu,-\mu)=\arctan\biggr\{\frac{f(\alpha;\mu)\sin[\chi(\alpha;\mu)]+\sin\theta\cos\varphi\; g(\alpha,-\alpha;\mu,-\mu)\sin[h(\alpha,-\alpha;\mu,-\mu)]}{f(\alpha;\mu)\cos[\chi(\alpha;\mu)]+\sin\theta\cos\varphi\; g(\alpha,-\alpha;\mu,-\mu)\cos[h(\alpha,-\alpha;\mu,-\mu)]}\biggr\}\nonumber \\
 &  & +\frac{1}{1+\sin\theta\cos\varphi}\biggr\{\omega_{1}\tau(\frac{1}{2}+\rho_{\alpha}^{2})+\omega_{2}\tau(\frac{1}{2}+\rho_{\mu}^{2})+\sin\theta\cos\varphi\exp[-2(\rho_{\alpha}^{2}+\rho_{\mu}^{2})][\omega_{1}\tau(\frac{1}{2}-\rho_{\alpha}^{2})+\omega_{2}\tau(\frac{1}{2}-\rho_{\mu}^{2})]\biggr\}.\label{eq:GeometricPhasePositiveNegative}
\end{eqnarray}
 \end{widetext}

\subsection{Discussions}

The following discussions are confined to Eq. \eqref{eq:GeometricPhasePositiveNegative}.
Namely we explore the properties of the geometric phase for the specific
type of entangled coherent state with $\beta=-\alpha$ and $\nu=-\mu$.

\emph{Case 1}. First we examine the effect of entanglement on the
geometric phase. Consider $\theta=0$ or $\theta=\pi$, and this leads
to vanishing entanglement, or product coherent states. Therefore,
the geometric phase is

\begin{eqnarray}
\gamma(\alpha;\mu) & = & \gamma(-\alpha;-\mu)\\
 & = & |\alpha|^{2}(\omega_{1}\tau-\sin\omega_{1}\tau)+|\mu|^{2}(\omega_{2}\tau-\sin\omega_{2}\tau),\nonumber
\end{eqnarray}
 It is clear that the geometric phase is the exact addition of geometric
phases \eqref{eq:GeometricPhaseOfOneParticle} acquired by either
subsystem. Our result shows that the geometric phase of any product
coherent state is equal to the sum of geometric phases acquired by
either subsystem. It is not difficult to find that only when $|\alpha|^{2}\rightarrow\infty$
and $|\mu|^{2}\rightarrow\infty$, the total phase vanishes. So under
the case of finite valued parameters, it is not necessary to study
the corresponding off-diagonal phase \cite{manini2000off,mukunda2001bargmann}.
Under the condition that $\theta=\pi/2$, the entanglement of this
state becomes maximal. Unlike the case of entangled spin pairs \cite{Sjoqvist2000geometric},
the dynamical phase still exists and $\langle\psi(0)|\psi(\tau)\rangle$
is not a real number, and hence the geometric phase \eqref{eq:GeometricPhasePositiveNegative}
also emerges.

\emph{Case}\textit{ 2}. Let us look at cyclic two-particle geometric
phase from Eq. \eqref{eq:GeometricPhasePositiveNegative}. When $\tau$
satisfies the following conditions
\[
\begin{cases}
\omega_{1}\tau= & 2\pi l_{1}\\
\omega_{2}\tau= & 2\pi l_{2}
\end{cases},
\]
 where $l_{1}$ and $l_{2}$ are integers, the corresponding wave
function takes the form
\begin{equation}
|\psi(\tau)\rangle=e^{-i(\pi l_{1}+\pi l_{2})}|\psi(0)\rangle,\label{eq:CyclicEvolution}
\end{equation}
 which is a cyclic evolution \cite{aharonov1987phase}. The we obtain
cyclic geometric phase for entangled coherent states, \begin{widetext}
\begin{eqnarray}
\begin{array}{lll}
\gamma^{c} & = & -(\pi l_{1}+\pi l_{2})+2\pi\frac{l_{1}(\frac{1}{2}+\rho_{\alpha}^{2})+l_{2}(\frac{1}{2}+\rho_{\mu}^{2})+\sin\theta\cos\varphi\exp[-2(\rho_{\alpha}^{2}+\rho_{\mu}^{2})][l_{1}(\frac{1}{2}-\rho_{\alpha}^{2})+l_{2}(\frac{1}{2}-\rho_{\mu}^{2})]}{1+\sin\theta\cos\varphi},\\
 & = & \gamma_{1}^{c}+\gamma_{2}^{c}
\end{array}\label{eq:CyclicGeometricPhase}
\end{eqnarray}
 \end{widetext}
where $\gamma_{1}^{c}$ is given in Eq. \eqref{eq:GeometricPhaseOfOneInPotentialCyclic} and $\gamma_{2}^{c}$ takes a similar form. Similar to the result of entangled spin pairs in
Ref. \cite{tong2003entangled}, when the system of entangled coherent
states undergoes cyclic evolutions \eqref{eq:CyclicEvolution}, the
total geometric phase is equal to the sum of all subsystem, no matter
whether entanglement exists or not.

\emph{Case}\textit{ 3}. We then consider how harmonic potential affects
the geometric phases. Assume that only particle $1$ is in the harmonic
potential, and the geometric phase of particle $1$ is found to be
\begin{widetext}
\begin{equation}
\begin{array}{c}
\gamma_{1}=\\
\arctan\{\frac{\exp[-\rho_{\alpha}^{2}(1-\cos\omega_{1}\tau)]\sin[-(\rho_{\alpha}^{2}\sin\omega_{1}\tau+\frac{1}{2}\omega_{1}\tau)]+\sin\theta\cos\varphi\exp[-\rho_{\alpha}^{2}(1+\cos\omega_{1}\tau)-2\rho_{\mu}^{2}]\sin(\rho_{\alpha}^{2}\sin\omega_{1}\tau-\frac{1}{2}\omega_{1}\tau)}{\exp[-\rho_{\alpha}^{2}(1-\cos\omega_{1}\tau)]\cos[-(\rho_{\alpha}^{2}\sin\omega_{1}\tau+\frac{1}{2}\omega_{1}\tau)]+\sin\theta\cos\varphi\exp[-\rho_{\alpha}^{2}(1+\cos\omega_{1}\tau)-2\rho_{\mu}^{2}]\cos(\rho_{\alpha}^{2}\sin\omega_{1}\tau-\frac{1}{2}\omega_{1}\tau)}\}-\delta_{1}
\end{array}\label{eq:GeometricPhaseOfOneInPotential}
\end{equation}
 via setting $\omega_{2}=0$, where the one-particle dynamical phase
$\delta_{1}$ is
\[
\delta_{1}=-\frac{\omega_{1}\tau(\frac{1}{2}+\rho_{\alpha}^{2})+\sin\theta\cos\varphi\exp[-2(\rho_{\alpha}^{2}+\rho_{\mu}^{2})][\omega_{1}\tau(\frac{1}{2}-\rho_{\alpha}^{2})]}{1+\sin\theta\cos\varphi}.
\]
 \end{widetext} Though particle $2$ does not experience harmonic
potential, it also has an influence on the geometric phase of particle
$1$ through the entanglement between the two particles. By setting
$\theta=0$ or $\pi$, the geometric phase \eqref{eq:GeometricPhaseOfOneInPotential}
reduces to Eq. \eqref{eq:GeometricPhaseOfOneParticle}. This is due
to the fact that the two particles are not entangled in this case.
When the system undertakes a cyclic evolution, i.e., $\omega_{1}\tau=2\pi l_{1}$,
we obtain the cyclic one-particle geometric phase which is \begin{widetext}
\begin{equation}
\gamma_{1}^{c}=-\pi l_{1}+\frac{2\pi}{1+\sin\theta\cos\varphi}\{l_{1}(\frac{1}{2}+\rho_{\alpha}^{2})+l_{1}(\frac{1}{2}-\rho_{\alpha}^{2})\sin\theta\cos\varphi\exp[-2(\rho_{\alpha}^{2}+\rho_{\mu}^{2})]\}.\label{eq:GeometricPhaseOfOneInPotentialCyclic}
\end{equation}
We also find two-particle dynamical phase as 
\begin{equation}
\begin{array}{lll}
\delta & = & -\frac{\omega_{1}\tau(\frac{1}{2}+\rho_{\alpha}^{2})+\omega_{2}\tau(\frac{1}{2}+\rho_{\mu}^{2})+\sin\theta\cos\varphi\exp[-2(\rho_{\alpha}^{2}+\rho_{\mu}^{2})][\omega_{1}\tau(\frac{1}{2}-\rho_{\alpha}^{2})+\omega_{2}\tau(\frac{1}{2}-\rho_{\mu}^{2})]}{1+\sin\theta\cos\varphi}\\
 & = & \delta_{1}+\delta_{2}
\end{array}.\label{eq:DynamicalPhaseOfPositiveNegative}
\end{equation}
 \end{widetext} It is reasonable to draw a conclusion that the total
dynamical phase is identical with sum of counterparts of subsystems
which is partially affected by the potential. It is worth to point
out that the result is in accordance with that for spin particles
as given in Ref. \cite{tong2003entangled}.

\section{Conclusion}

\label{sec:discussion}

We have derived an elegant expression of geometric phase for coherent
states. The non-cyclic geometric phases for entangled coherent states and
some of their properties are investigated. We explore the influences
of entanglement and potential on the geometric phases. Our results
show that when entanglement vanishes, the geometric phase of product
state is equal to the sum of counterparts of individual particles.
In the case of maximally entangled coherent states, the dynamical
phase and geometric phase do not disappear, and the result is different
from that for entangled spin pairs. We also explore the property of
cyclic two-particle geometric phases. It is found that the cyclic
geometric phase is identical with addition of counterparts of subsystems
no matter whether entanglement vanishes or not. Our findings are consistent
with the results for entangled spin pairs. If only one particle is
affected by harmonic potential, both un-cyclic and cyclic one-particle
geometric phases are affected by the other particle due to the existence
of entanglement between the two particles. It is worth to mention
that two-particle dynamical phase is not equal to the addition
of counterparts of particles 1 and 2, while it can be expressed in
terms of $\delta_{1}+\delta_{2}$, where $\delta_{1}/\delta_{2}$
is cyclic dynamical phase of particle $1/2$ experiencing harmonic
potential while no potential acting on particle $2/1$ respectively.

\section{Acknowledgements}

D.B.Y. acknowledges Rui Zhang for her helpful discussions. J.L.C.
is supported by National Basic Research Program (973 Program) of China
under Grant No. 2012CB921900 and NSF of China (Grant Nos. 10975075
and 11175089). This work is also partly supported by National Research
Foundation and Ministry of Education , Singapore (Grant No. WBS: R-710-000-008-271).

\end{document}